\documentclass{article}
\usepackage{amssymb}
\usepackage{graphicx}

\newtheorem{theorem}{Theorem}

\newtheorem{proposition}[theorem]{Proposition}

\input{tcilatex}
\begin{document}

\title{Uniform Conversions of Operating Points and Characteristics of
Compressor}
\author{L. A. Ostromuhov \thanks{\textit{Dr. L. A. Ostromuhov, Wingas
Transport GmbH, Baumbachstr. 1, 34119 Kassel, Germany. E-mail:
leonid.ostromuhov@wingas-transport.de}}}
\date{}
\maketitle

\begin{abstract}
In the paper, some aspects of the polytropic analysis are developed that
concerned with various processes changing the thermodynamic state of flow of
a real fluid and with reduction of these processes to processes having a
given temperature and pressure of a given real mixture at the inlet. It is
shown that all parameters of the process can be converted under condition of
full similarity of flow that is formulated in the paper. An operating point
of a compressor represents such process. It is to emphasize that parameters
of the reduced point include not only volume flow, speed and polytropic
head, for which a requirement of similarity of flow at inlet is sufficient,
but also polytropic exponent, polytropic efficiency, outlet pressure and
outlet temperature. This allows a conversion of all compressor
characteristics. A method for that is described.
\end{abstract}

\tableofcontents

\section{Introduction}

In the paper, some aspects of the polytropic analysis are developed that
concerned with various processes changing the thermodynamic state of flow of
a real fluid and reduction of these processes to processes having a given
temperature and pressure of a given real mixture at the inlet. It is shown
that all parameters of the process can be converted under condition of full
similarity of flow that is formulated in the paper. So dynamic similitude is
achieved. Every such conversion is defined after mechanical and
thermodynamic properties of both the process to be converted and the given
state of mixture at inlet of the reduced process. The conversions are
unique, reversible, and they construct a group.

An operating point of a compressor represents such process. Let be given
some fixed reference gas composition and reference pressure and temperature.
When they are considered at inlet, they are called the reference inlet
conditions. Let a compressor operating point has some different gas
composition and different inlet pressure and temperature. In many
applications and for comparison, the different operating points must be
reduced to the reference inlet conditions. In literature and practice
before, by the test evaluation, by the conversion of test results to
guarantee conditions and comparison with guaranted values, there was
considered just well-known fan laws and conditions of similarity of flow at
inlet. They allow to convert volume flow, speed and polytropic head from the
operating conditions to the reference inlet conditions. However, they are
not sufficient to convert the polytropic exponent, polytropic efficiency,
outlet pressure and outlet temperature.

In this paper, it is shown that for any compressor operating point, a
condition of similarity of flow both at inlet and at outlet defines uniquely
a reduced point that corresponds to the reference inlet conditions so that
there is dynamic similitude between the operating point and the reduced
point. This allows to construct a conversion of all compressor
characteristics from one given inlet conditions to another ones. A method
for that is described. The results can be generalized achieving dynamic
similitude between operating point of the compressor and the reduced point
of a geometrically similar machine. The condition of similarity of flow both
at inlet and at outlet is named as condition of full similarity of flow. It
is to emphasize that parameters of the reduced point include not only volume
flow, speed and polytropic head, for which a requirement of similarity of
flow at inlet is sufficient, but also polytropic exponent, polytropic
efficiency, outlet pressure and outlet temperature.

\subsection{Problem}

Let us consider an uncooled multistage centrifugal compressor with one
suction as inlet and one discharge as outlet without side-flow. We suppose
that hydraulic losses in diffusors and between the compressor stages, the
dependence of the losses on the gas composition, roughness and Reynolds
number, and their influence on performance and characteristics of compressor
are treated by a known way e.g. as in \cite{ICAAMC}.

In the polytropic analysis of uncooled centrifugal compressors, J. M.
Schultz \cite{Schultz} defined the unique path of compression between two
thermodynamic states of gas as a path with the given constant efficiency.
This path plays a central role in test evaluation and design of compressors.

Let be given some fixed reference gas composition and reference pressure and
temperature. When they are considered at inlet, they are called the
reference inlet conditions. Let a compressor operating point has some
different gas composition and different inlet pressure and temperature. In
many applications and for comparison, the different operating points must be
reduced to the reference inlet conditions.

An operating point of the compressor is a point in a set of its parameters
such as its inlet and outlet and intermediate parameters including gas
composition. So the operating point includes two thermodynamic states of
gas, namely states at the inlet and outlet of the compressor, and a path
between these states. Its operating inlet conditions, consisting of the
operating gas composition and operating gas pressure and temperature at
inlet, may be different from the reference ones.

If the operating point undergoes a transformation under the action of
certain rules, then the conversion result is called the converted point.

In particularly, if the conversion rules include the reference inlet
conditions, then the converted point is called as the reduced point or as
the reference point of compressor.

In many applications, it is necessary to reduce any compressor point from
its operating conditions to the reference conditions i.e. to convert the
operating point of compressor to the reduced point of a similar machine so
that similitude is achieved. It means that an operating value of every
parameter that has been given, measured or calculated for the operating gas
composition and the operating inlet conditions must be converted to its
reduced value that corresponds to the reference gas composition and the
reference inlet conditions so that there is dynamic similitude between the
operating and reduced values.

It is also useful to learn how to convert compressor characteristics given
for the reference gas composition and the reference inlet conditions to
operating ones and vice versa.

P. 2, 3 contain mostly the well-known terms and approaches used before by
guarantee tests and performance monitoring. It is shown that these
approaches must be completed.

The fan laws are the result of the conditions of similarity of flow at
inlet. They allow to convert volume flow, speed and polytropic head from the
operating conditions to the reference conditions \cite{Schultz}, \cite{VDI
2045 p2}. The state of art of test evaluation, conversion of test results to
guarantee conditions and comparison with guaranted values is described in 
\cite{VDI 2045 p1}. The main objective values of conversion are (i)
effective intake volume flow, (ii) pressure ratio, (iii) power at the
coupling.

However, the fan laws as well as the results described in \cite{Schultz}, 
\cite{VDI 2045 p1}, \cite{VDI 2045 p2} are not sufficient to convert
polytropic exponent, polytropic efficiency, outlet pressure and outlet
temperature. For these purposes, there were attempts to use additional
suggestions such as

- invariance of polytropic efficiency,

- a certain rool of proportionality between the operating and reduced
isentropic exponents taken for inlet states and the operating and reduced
polytropic exponents.

In the paper, it is shown and numerical results are provided that the last
two suggestions are inconsistent and contradict each other. None of them
lead to dynamic similitude.

In p. 4 -- 6 of this paper, it is shown that for any compressor operating
point, a suggestion of similarity of flow both at inlet and outlet defines
uniquely a reduced point that corresponds to the reference inlet conditions
so that there is dynamic similitude between the operating point and the
reduced point. This allows to construct a conversion of all compressor
characteristics from one given inlet conditions to another ones. The results
can be generalized achieving dynamic similitude between operating point of
the compressor and reduced point of a geometrically similar machine. The
suggestion of similarity of flow both at inlet and outlet is named as
suggestion or condition of full similarity of flow. It is to emphasize that
this suggestion allows to determine all parameters of reduced point include
not only volume flow, speed and polytropic head, for which the requirement
of similarity of flow at inlet is sufficient, but also polytropic exponent,
polytropic efficiency, outlet pressure and outlet temperature.

\subsection{Applications using the coversion problem}

The coversion problem above and the condition of full similarity of flow
play a role in the following applications working at the following machine
configurations:

1. Applications for a single machine:

-- Performance Monitoring

-- Simulation of operating points

-- Characteristics diagnostics, i.e. Determination of actual characteristics
of machine

-- Model Diagnostics, i.e. Determination of actual parameters of models

-- Design

2. Applications for a shop of machines consisting of parallel machines of
the same type:

-- Simulation of operating points

3. Applications for a station consisting of several machines:

-- Simulation of operating points

-- Load sharing

-- Optimal load sharing and control in real time

4. Applications for a gas network:

-- Steady state simulation

-- Dynamic simulation

-- Load sharing

-- Optimal load sharing and optimization of the network state under the
various economic and operating objective functions

-- Optimal control.

All these applications except for dynamic simulation and optimal control are
realized in the systems ACCORD, Graphicord, Compressor-Visualizer and
Compressor-Characterizer as their components.

\subsection{Specific applications and their implementation}

The work grew out from the applications that autor set, defined, carried out
or participated. Some of these applications are:

-- ACCORD using for steady state simulation, optimal load sharing and
optimization of gas networks of any structure;

- Graphicord using for graphical user interface; data base and spread sheet
support; scenario management; management of the third party programs;
connection with the geographic information system; interface and data
exchange with the distributed control systems;.

-- Compressor-Visualizer using for performance monitoring and simulation of
operating points of compressors, turbines, and other drives;

-- Compressor-Characterizer using for performance monitoring,
characteristics diagnostics, and model diagnostics of compressors, turbines,
and other drives;

-- Compass using for performance monitoring, characteristics diagnostics,
and simulation of operating points of compressors and turbines.

Historically, ACCORD, Compressor-Visualizer and Compressor- Characterizer
were developed at first. Then features mentioned above, which concerned with
performance monitoring, simulation and characteristics diagnostics of
compressors and turbines, were developed in Compass, Co. Bruel and Kjaer, in
collaboration with Co. Wingas Transport additionally to vibrodiagnostics
that Compass contained before. From the side of Bruel and Kjaer, the work
was carried out by E. Andreassen. Compressor-Visualizer and Compressor-
Characterizer were developed by A. I. Gutnikov and author.

The author is glad to thank E. Andreassen, A. I. Gutnikov, A. E. Lapitsky
and his colleagues at Wingas Transport for fruitful discussions and
cooperation.

\section{Basic concepts of compressor tests}

\subsection{The main parameters of compressor}

Let us describe at first a minimal set of parameters allowing a description
of the compressor. The Mach and Reynolds numbers are added to the following
list of parameters to ensure the use of flow similarity theory.

1. Parameters of the compressor inlet

-- $V_{1}^{\prime }$ -- Volumetric flow

-- $P_{1}$ -- Pressure

-- $T_{1}$ -- Temperature

-- $Z_{1}$ -- Compressibility factor

-- $k_{1}$ -- Isentropic exponent

-- $Ma_{1}$ -- Mach number

-- $Re_{1}$ -- Reynolds number

2. Parameters of the compressor outlet

-- $V_{2}^{\prime }$ -- Volumetric flow

-- $P_{2}$ -- Pressure

-- $T_{2}$ -- Temperature

-- $Z_{2}$ -- Compressibility factor

-- $k_{2}$ -- Isentropic exponent

-- $Ma_{2}$ -- Mach number

-- $Re_{2}$ -- Reynolds number

3. Generic compressor parameters

-- $Mix$ -- Gas composition

-- $V_{n}^{\prime }$ -- Volumetric flow at STP (Standard condition for
temperature and pressure)

-- $N$ -- Speed of rotation

-- $H_{p}\ $-- Polytropic head, mass-specific

-- $n$ -- Polytropic exponent

-- $S=P_{2}/P_{1}$ -- Pressure ratio

-- $E_{p}\ $-- Polytropic efficiency

-- $W$ -- Compressor power, required

\subsection{Dimensionless characteristic numbers of compressor used for
conditions of similarity of flows of two fluids}

Like in \cite{VDI 2045 p2}, pages 35 - 37, we list the following
dimensionless characteristic numbers of compressor using as criteria of
similarity. Invariance, here -- identity, of characteristic numbers means
similarity and allows conversion of according parameters from values known
for one conditions to values valid for another conditions.

\subsubsection{Flow coefficient $\protect\varphi $}

\begin{equation}
\varphi =\frac{V^{\prime }}{\pi D^{2}u/4}  \label{def_fi}
\end{equation}%
where $u$ is the tip speed of the impeller and $\pi D^{2}/4$ is an
impereller cross-section area.

Suppose that the fluid flow does not come off from the wheel or its blades.
Then the tip speed is

\begin{equation}
u=\pi DN  \label{u_on_ND}
\end{equation}

In multi-stage machines, the flow $V^{\prime },$ the tip speed $u,$ the
impereller diameter and cross-section area can be related to the first stage

\begin{equation}
\varphi _{1}=\frac{V_{1}^{\prime }}{\pi D_{1}^{2}u_{1}/4}
\end{equation}%
or to the last stage for the outlet conditions.

\subsubsection{Head coefficient i.e. process work coefficient $\protect\psi $%
}

\begin{equation}
\psi =\frac{Y}{u^{2}/2}  \label{def_psi}
\end{equation}

In multi-stage machines, compression work $Y$ can be related to the entire
compressor while kinetic energy and tip speed can be related to the first
stage

\begin{equation}
\psi _{1}=\frac{Y}{u_{1}^{2}/2}
\end{equation}%
or to the last stage for the outlet conditions.

\subsubsection{Tip speed Mach number $Ma.$}

For multi-stage machines, a tip speed Mach number is

\begin{equation}
Ma=\frac{u}{a}  \label{def_Ma}
\end{equation}%
i.e. the tip speed of considering stage divided by the speed of sound as it
is in fluid, which has a considering composition $Mix$ and whose state
corresponds to specific physical conditions of temperature and pressure,

The speed of sound can be calculated as

\begin{equation}
a=\sqrt{kR_{g}TZ}  \label{def_v_sonic}
\end{equation}

\begin{equation}
R_{g}=\frac{R_{0}}{\mu }
\end{equation}

We will call the tip speed Mach number as Mach number often for simplicity.
So the inlet Mach number is

\begin{equation}
Ma_{1}=\frac{u_{1}}{a_{1}}=\frac{\pi D_{1}N_{1}}{\sqrt{k_{1}R_{g}T_{1}Z_{1}}}
\label{def_Ma1}
\end{equation}%
and the outlet Mach number is

\begin{equation}
Ma_{2}=\frac{u_{2}}{a_{2}}=\frac{\pi D_{2}N_{2}}{\sqrt{k_{2}R_{g}T_{2}Z_{2}}}
\label{def_Ma2}
\end{equation}

\subsubsection{Tip Reynolds number $Re$}

Tip Reynolds number is

\begin{equation}
Re=\frac{ub}{\nu }
\end{equation}

So the inlet Reynolds number is 
\begin{equation}
Re_{1}=\frac{u_{1}b_{1}}{\nu _{1}}=\frac{\pi D_{1}N_{1}b_{1}}{\nu _{1}}
\end{equation}%
and the outlet Reynolds number is

\begin{equation}
Re_{2}=\frac{u_{2}b_{2}}{\nu _{2}}=\frac{\pi D_{2}N_{2}b_{2}}{\nu _{2}}
\end{equation}%
where kinematic viscosity $\nu _{1}$ is referred to the inlet state of gas
at the first stage and $\nu _{2}$ is referred to the outlet state of gas at
the last stage.

We will consider below just the case of 
\begin{equation}
N_{1}=N_{2},
\end{equation}%
when the first and last stages have identical rotation speed, e.g. when the
stages are located on the same shaft.

\subsection{Fan laws}

Fan laws relates to the definition of the flow coefficient $\varphi $ and
the head coefficient $\psi .$ The fan laws mean that the volume flow $%
V^{\prime }$ increases linearly with speed and the polytropic head $H_{p}$
increases quadratically. In the form of tip speed, they means

\begin{equation}
V^{\prime }=\varphi ^{\prime }N  \label{Fan_law_1st}
\end{equation}

\begin{equation}
H_{p}=\psi ^{\prime }N^{2}  \label{Fan_law_2nd}
\end{equation}

For the suction volume flow and for the first stage

\begin{equation}
V_{1}^{\prime }=\varphi _{1}^{\prime }N
\end{equation}

\begin{equation}
H_{p}=\psi _{1}^{\prime }N^{2}  \label{Fan_law_2nd_shtrix}
\end{equation}

Conditions of compressor similarity include invariance of the both
coefficients of flow and process work. The invariance of the fan laws for
two conditions $c$ and $a$ means that their $\varphi ,\psi $ are the same.
The invariance can be considered relative to the speed $u$ and $N$ in the
form of $\varphi _{c}=\varphi _{a},$ $\psi _{c}=\psi _{a}$ as well as $%
\varphi _{c}^{\prime }=\varphi _{a}^{\prime },$ $\psi _{c}^{\prime }=\psi
_{a}^{\prime }$ respectively.

For simplicity, we will write $\varphi ,\psi $ instead of $\varphi ^{\prime
},\psi ^{\prime },$ where it is not controversial.

\section{State of the art: Determination of actual parameters and their
partial conversion under laws of similarity of flow at inlet}

\subsection{Use of equation of state and thermodynamic properties of real
gases}

Let us suppose that a form of equation of state for real gases is chosen for
using in calculations or that tables and diagrams for thermodynamic
properties are chosen. So we can assume that we possess the computing
functions for all the thermodynamic quantities that are functions of gas
composition $Mix$ of thermodynamic state $\left( T,P,Mix\right) $ e.g.

\begin{equation}
R_{g}=R\left( Mix\right) =\frac{R_{0}}{\mu \left( Mix\right) }
\label{f_R_gas}
\end{equation}

\begin{equation}
Z=T\left( T,P,Mix\right)  \label{f_z}
\end{equation}

\begin{equation}
k=k\left( T,P,Mix\right)  \label{f_k}
\end{equation}

\begin{equation}
H=H\left( T,P,Mix\right)  \label{f_h}
\end{equation}

The calculations below are valid for any compressor stage and for
multi-stage compressor due to use of polytropic path like in \cite{Schultz}.

\subsection{Actual Performance}

From the main parameters of compressor listed above, let be given or
measured $Mix_{a},$ $V_{1,_{a}}^{\prime },$ $P_{1,_{a}},$ $T_{1,_{a}},$ $%
P_{2,_{a}},$ $T_{2,_{a}},$ and $N_{a}.$ On the base of these values, we can
calculate the thermodynamic properties at inlet and outlet as follows.

The pressure ratio is 
\begin{equation}
S_{a}=\frac{P_{2,_{a}}}{P_{1,_{a}}}  \label{AP__S_a}
\end{equation}%
The compressibility factors $Z_{1,a},Z_{2,a}$ are calculated due to (\ref%
{f_z}). The polytropic exponent is 
\begin{equation}
n_{p,a}=\frac{\ln S_{a}}{\ln S_{a}-\ln \left( \frac{Z_{2,a}T_{2,a}}{%
Z_{1,a}T_{1,a}}\right) }  \label{AP__n_p_a}
\end{equation}%
With 
\begin{equation}
A_{n_{p,a}}=\frac{n_{p,a}-1}{n_{p,a}}
\end{equation}%
the polytropic head is 
\begin{equation}
H_{p,a}=\frac{1}{A_{n_{p,a}}}\left( S_{a}^{A_{n_{p,a}}}-1\right)
R_{a}T_{1,a}Z_{1,a}  \label{AP__H_p_a}
\end{equation}%
The enthalpies at inlet and outlet $H_{1,a},H_{2,a}$ are calculated due to (%
\ref{f_h}). Ratio of the polytropic head with the enthalpy rise gives the
polytropic efficiency: 
\begin{equation}
E_{p,a}=\frac{H_{p,a}}{H_{2,a}-H_{1,a}}
\end{equation}

A compressor power corresponds to the work expended on the gas during
compression. The required compression power in shaft is

\begin{equation}
W_{a}=\frac{1}{E_{p,a}}H_{p,a}m_{a}^{\prime }=\frac{1}{E_{p,a}}\frac{1}{%
A_{n_{p,a}}}\left( S_{a}^{A_{n_{p,a}}}-1\right) P_{1,_{a}}V_{1,_{a}}^{\prime
}  \label{W_on_H}
\end{equation}

\subsection{Identity of Mach numbers and flow coefficients at inlet and its
using for conversion of operating point}

Let be given compositions of 2 fluids $Mix_{a}$ and $Mix_{r}$ and 2
according inlet conditions $\left( T_{1,a},P_{1,a}\right) $ and $\left(
T_{1,r},P_{1,r}\right) $. For these fluids and states, gas properties such
as $k,Z,\rho $ can be calculated and denoted by the corresponding indexes.

For the fluid $Mix_{a}$ and inlet conditions $\left( T_{a},P_{a}\right) ,$
let there is given an operating point $a.$ It means that we know all its
data as measured or calculated values such as the inlet flow rate $%
V_{1,a}^{\prime },$\ the tip speed $N_{1,a},$\ the polytropic head $H_{p,a},$
the polytropic efficiency $E_{p,a},$ the outlet temperature $T_{2,a}$ and
pressure $P_{2,a},$ the required power and so on.

Having point $a,$ let find now, which parameters can be found for a point $c$
that has the reference composition $Mix_{c}=Mix_{r}$ and the reference inlet
conditions $\left( T_{1,c},P_{1,c}\right) =\left( T_{1,r},P_{1,r}\right) ,$
if we suppose that points $c$ and $a$ have identical Mach numbers and flow
coefficients at inlet of the compressor:

\begin{equation}
Ma_{1,c}=Ma_{1,a}.  \label{Ma1c_eq_Ma1a}
\end{equation}

\begin{equation}
\varphi _{1,c}=\varphi _{1,a}.  \label{fi1c_eq_fi1a}
\end{equation}%
Remember that we consider a compressor constructed so that $N_{1}=N_{2}=N$
i.e. the first and last stages have identical rotation speed.

Then due to (\ref{Ma1c_eq_Ma1a}), (\ref{fi1c_eq_fi1a}), (\ref{def_Ma}), (\ref%
{def_v_sonic}), (\ref{u_on_ND}), (\ref{Fan_law_1st})

\begin{equation}
\frac{N_{a}}{\sqrt{k_{1,a}R_{a}T_{1,a}Z_{1,a}}}=\frac{N_{c}}{\sqrt{%
k_{1,r}R_{r}T_{1,r}Z_{1,r}}}  \label{conj3-n}
\end{equation}%
and

\begin{equation}
\frac{V_{1,a}^{\prime }}{\sqrt{k_{1,a}R_{a}T_{1,a}Z_{1,a}}}=\frac{%
V_{1,c}^{\prime }}{\sqrt{k_{1,r}R_{r}T_{1,r}Z_{1,r}}}  \label{conj3-v}
\end{equation}

With a conversion factor defining as

\begin{equation}
C_{ca}=\frac{a_{1,r}}{a_{1,a}}=\sqrt{\frac{k_{1,r}R_{r}T_{1,r}Z_{1,r}}{%
k_{1,a}R_{a}T_{1,a}Z_{1,a}}}  \label{__C_F}
\end{equation}%
we get

\begin{equation}
N_{c}=C_{ca}N_{a}  \label{conv_N}
\end{equation}

\begin{equation}
V_{1,c}^{\prime }=C_{ca}V_{1,a}^{\prime }  \label{conv_V}
\end{equation}%
and the first fan law at inlet that is identity of the flow coefficients $%
\varphi ^{\prime }$

\begin{equation}
\varphi _{1,c}^{\prime }=\varphi _{1,a}^{\prime }
\end{equation}%
where $\varphi _{1}^{\prime }$ is taken in the form $\varphi _{1}^{\prime
}=V_{1}^{\prime }/N_{1}.$

These are conditions that concern speed and flow correspondingly and that
are like the first fan law.

\subsection{Identity of the process work coefficient $\protect\psi $ in the
second fan law and its using for conversion of operating point}

For conversion of the operating point, we suppose identity of $\psi $ in the
second fan law using in the form

\begin{equation}
H_{p}=\psi ^{\prime }N^{2}.
\end{equation}%
Identity of $\psi $ means that

\begin{equation}
\psi _{c}^{\prime }=\psi _{a}^{\prime }  \label{PSIc_eq_PSIa}
\end{equation}%
Combining (\ref{PSIc_eq_PSIa}) with (\ref{conv_N}), we get

\begin{equation}
H_{p,c}=\left( \frac{N_{c}}{N_{a}}\right) ^{2}H_{p,a}=C_{ca}^{2}H_{p,a}
\label{conv_Hp}
\end{equation}

We can conclude that the use of identity of the Mach numbers $Ma_{1}$ at
inlet (\ref{Ma1c_eq_Ma1a}), the flow coefficients at inlet $\varphi _{1}$ (%
\ref{fi1c_eq_fi1a}) and the process work coefficients $\psi $ (\ref%
{PSIc_eq_PSIa}) ensures conversion of the inlet flow $V_{1}^{\prime }$, the
speed $N$ and the polytropic process work $H_{p}$. Their converted three
parameters $\left( V_{1,c}^{\prime },H_{p,c},N_{c}\right) $ represent a
point in the same space, where a diagramm of compressor characteristics is
often given by manufacturer for the reference gas composition and the
reference inlet conditions. For any pair $\left( V_{1,c}^{\prime
},N_{c}\right) $ of inlet flow and speed, a value of polytropic head $%
H_{p,e} $ can be read from these manufacturer characteristics. The read
value is called as expected value. This allows a comparison between $%
H_{p,c}, $ the operating value converted to the reference conditions, and $%
H_{p,e},$ the expected value given for these conditions by manufacturer or
vendor. No other value can be converted and compared.

\section{Full conversion of compressor operating point under suggestion of
similarity of flows both at inlet and at outlet}

We see that if laws of similarity of flow at inlet are used to conversion of
parameters of the compressor operating point and no additional suggestion is
used then we lack conversion of outlet temperature and pressure, polytropic
exponent, polytropic efficiency, required power and so on. Determination of
the reduced outlet state $(T_{2,c},P_{2,c})$ and the reduced polytropic
exponent $n_{p,c}$ concerns to identification of the reduced polytropic
process of the reference gas. We need to find these values so that

\begin{equation}
H_{p,c}=\frac{1}{A_{n_{p,c}}}\left( S_{c}^{A_{n_{p,c}}}-1\right)
R_{r}T_{1,r}Z_{1,r}  \label{def_H_p_c}
\end{equation}%
where $S_{c}=P_{2,c}/P_{1,r}.$

\subsection{The main proposition}

There are given i.e. known

1) $Mix_{r}=$ composition of the reference gas mixture

2) $(T_{1,r},P_{1,r})=$ the reference inlet conditions i.e. temperature and
pressure at inlet of the reference gas

3) $Mix_{a}=$ composition components of operating gas mixture

4) $(T_{1,a},P_{1,a})=$ operating inlet conditions of operating gas

5) $(T_{2,a},P_{2,a})=$ operating outlet conditions of operating gas.

The values 3 - 5 relate to an operating polytropic process of operating gas.

The condition of similarity suggested at inlet includes an invariance of the
inlet Mach numbers $Ma_{1}$ (\ref{Ma1c_eq_Ma1a}) and the inlet flow
coefficients $\varphi _{1}$ (\ref{fi1c_eq_fi1a}). It determines an inlet
volume of the reference gas. The third suggested conditions of similarity is
an invariance of the process work coefficient $\psi $ (\ref{PSIc_eq_PSIa}).
It determines a polytropic head of the reference gas.

The outlet conditions of the reference gas are unknown. So the polytropic
process of the reference gas is unknown and must be found.

The suggestion that not only the inlet flows of the operating and reference
gases are similar but also the outlet flows are similar can be written as

\begin{eqnarray}
reference\ inlet\ conditions &:&  \label{FS_ref} \\
Mix_{c} &=&Mix_{r},  \label{FS_ref_Mix} \\
(T,P)_{1,c} &=&(T,P)_{1,r}  \label{FS_ref_TP}
\end{eqnarray}

\begin{equation}
Ma_{1,c}=Ma_{1,a}.  \label{FS_Ma1c_eq_Ma1a}
\end{equation}

\begin{equation}
Ma_{2,c}=Ma_{2,a}  \label{Ma2c_eq_Ma2a}
\end{equation}

\begin{equation}
Re_{1,c}=Re_{1,a}  \label{FS_Re1c_eq_Re1a}
\end{equation}

\begin{equation}
Re_{2,c}=Re_{2,a}  \label{FS_Re2c_eq_Re2a}
\end{equation}

\begin{equation}
\varphi _{1,c}=\varphi _{1,a}  \label{FS_fi1c_eq_fi1a}
\end{equation}

\begin{equation}
\varphi _{2,c}=\varphi _{2,a}  \label{fi2c_eq_fi2a}
\end{equation}

\begin{equation}
\psi _{c}=\psi _{a}  \label{FS_PSIc_eq_PSIa}
\end{equation}

The conditions of invariant Reynolds numbers are not used below because they
can be treated by a known way e.g. as in \cite{ICAAMC}. We do not discuss
here, if this way is sufficient.

\begin{proposition}
The suggestion of similarity of flows both at inlet and at outlet allows
conversion of all parameters of the compressor operating point to the
reference inlet conditions consisting of the reference gas composition and
the reference values of the inlet temperature and pressure so that there is
dynamic similitude between the operating and reduced values.
\end{proposition}

To show how the conversion is constructed we will determine the outlet
conditions of the reference gas $T_{2c},$ $P_{2c,}$ and the polytropic
exponent $n_{p,c}$ describing its polytropic path, if we suppose that not
only the inlet flows of the operating and reference gas are similar but also
their outlet flows are similar.

Remember that we consider below just the cases of

(i) the same uncooled machine;

(ii) $N_{1}=N_{2}=N,$ when the first and last stages have identical rotation
speed e.g. when they located on the same shaft. This case is applied to the
same machine.

At inlet, the reference value $r$ and the converted value $c$ relate to the
same value. Therefore and for uniformity, the reference gas $Mix_{r},$ its
reference inlet temperature $T_{1,r}$ and pressure $P_{1,r},$ and its
thermodynamic properties at inlet will be sometimes below denoted by index $%
c $ instead of $r.$

\subsection{Effects of similarity of flows both at inlet and at outlet}

Similarity of flows both at inlet (\ref{Ma1c_eq_Ma1a}), (\ref{fi1c_eq_fi1a})
and at outlet (\ref{Ma2c_eq_Ma2a}), (\ref{fi2c_eq_fi2a}) means due to (\ref%
{def_Ma}, \ref{def_v_sonic}, \ref{u_on_ND}, \ref{Fan_law_1st}, \ref{def_Ma1}%
, \ref{def_Ma2} ) that

\begin{equation}
\frac{N_{jc}}{N_{ja}}=\frac{a_{jc}}{a_{ja}},\quad j=1,2,
\end{equation}

\begin{equation}
\frac{V_{jc}^{\prime }}{V_{ja}^{\prime }}=\frac{a_{jc}}{a_{ja}},\quad j=1,2,
\end{equation}

Because of

\begin{equation}
N_{1a}=N_{2a}=N_{a}
\end{equation}

\begin{equation}
N_{1c}=N_{2c}=N_{c}
\end{equation}

we get

\begin{equation}
\frac{N_{c}}{N_{a}}=C_{ca}=\frac{a_{1c}}{a_{1a}}=\frac{\sqrt{%
k_{1c}R_{c}T_{1c}Z_{1c}}}{\sqrt{k_{1a}R_{a}T_{1a}Z_{1a}}}=\frac{a_{2c}}{%
a_{2a}}=\frac{\sqrt{k_{2a}R_{a}T_{2a}Z_{2a}}}{\sqrt{k_{2c}R_{c}T_{2c}Z_{2c}}}
\end{equation}

\begin{equation}
\frac{V_{1c}^{\prime }}{V_{1a}^{\prime }}=\frac{V_{2c}^{\prime }}{%
V_{2a}^{\prime }}=\frac{N_{c}}{N_{a}}  \label{__V_12}
\end{equation}

\subsection{Effect on formulas for polytropic process}

For polytropic process 
\begin{equation}
PV^{n}=Const
\end{equation}%
so 
\begin{equation}
\left( PV^{n}\right) _{1}=\left( PV^{n}\right) _{2}
\end{equation}%
Hence, due to definition of $Z$ as compressibility factor for any equation
of state, 
\begin{equation}
\frac{T_{2}Z_{2}}{T_{1}Z_{1}}=\left( \frac{P_{2}}{P_{1}}\right) ^{\frac{n-1}{%
n}}=\left( \frac{V_{1}}{V_{2}}\right) ^{n-1}
\end{equation}

Writing these formulas for polytropic process of the operating gas and then
for corresponding polytropic process of the reference gas, due to (\ref%
{__V_12}) 
\begin{equation}
\frac{V_{1c}^{\prime }}{V_{2c}^{\prime }}=\frac{V_{1a}^{\prime }}{%
V_{2a}^{\prime }}
\end{equation}%
we get 2 equations 
\begin{equation}
\left( \frac{P_{2c}}{P_{1c}}\right) ^{\frac{1}{n_{p,c}}}=\frac{%
V_{1a}^{\prime }}{V_{2a}^{\prime }}  \label{__fe_P_V}
\end{equation}%
\begin{equation}
\left( \frac{T_{2c}Z_{2c}}{T_{1c}Z_{1c}}\right) ^{\frac{1}{n_{p,c}-1}}=\frac{%
V_{1a}^{\prime }}{V_{2a}^{\prime }}  \label{__fe_T_V}
\end{equation}%
where right sides $\frac{V_{1a}^{\prime }}{V_{2a}^{\prime }},P_{1c},T_{1c},$
and $Z_{1c}=Z(T_{1c},P_{1c})$ are given. With a chosen equation of state for
the reference gas mixture, after which compressibility factor $Z$ can be
calculated, we have%
\begin{equation}
Z_{2c}=Z(T_{2c},P_{2c}).  \label{__fe_Z}
\end{equation}%
So 3 equations (\ref{__fe_P_V}, \ref{__fe_T_V}, \ref{__fe_Z}) for 4 unknown
variables $T_{2c},$ $P_{2c},$ $Z_{2c},$ and $n_{p,c}$ are constructed.

\subsection{Effect on formulas for polytropic head}

The forth equation follows from the second fan law (\ref{conv_Hp}) and the
generic formula for the polytropic head (\ref{def_H_p_c}). From (\ref%
{conv_Hp}), (\ref{__C_F}), (\ref{AP__H_p_a}, \ref{def_H_p_c}) 
\begin{equation}
\frac{1}{A_{n_{p,c}}}\left( \left( \frac{P_{2c}}{P_{1c}}\right)
^{A_{n_{p,c}}}-1\right) =\frac{C_{ca}^{2}}{R_{c}T_{1c}Z_{1c}}H_{p,a}
\label{__fe_Hp}
\end{equation}

Here $P_{2c}$ in $S_{c}$ and $n_{p,c}$ in $A_{n_{p,c}}$ are unknown, all
other parameters are given.

\subsection{System of equations for the polytropic exponent and the outlet
values of temperature, pressure, and compressibility factor}

As result, we get a system of 4 equations (\ref{__fe_P_V}, \ref{__fe_T_V}, %
\ref{__fe_Z}, \ref{__fe_Hp}) with $A_{n_{p,c}}=\frac{n_{p,c}-1}{n_{p,c}},$
where right sides $\frac{V_{1a}^{\prime }}{V_{2a}^{\prime }},P_{1c},T_{1c},$
and $Z_{1c}=Z(T_{1c},P_{1c}),H_{p,a},C_{ca}$ are given. We like to find $%
T_{2c},P_{2c},n_{p,c},Z_{2c}.$

Introducing as the calculated given parameters 
\begin{equation}
D_{F}=D_{c1,a12}=\frac{C_{ca}^{2}}{R_{c}Z_{1c}T_{1c}}H_{p,a}
\end{equation}

\begin{equation}
V_{F}=V_{Fa}=\frac{V_{1a}^{\prime }}{V_{2a}^{\prime }}
\end{equation}%
and substituting (\ref{__fe_P_V}) in (\ref{__fe_Hp}), we get 
\begin{equation}
\frac{1}{A_{n_{p,c}}}\left\{ \left[ \left( V_{F}\right) ^{n_{p,c}}\right] ^{%
\frac{n_{p,c}-1}{n_{p,c}}}-1\right\} =D_{F}
\end{equation}%
i.e.%
\begin{equation}
\frac{n_{p,c}}{n_{p,c}-1}\left\{ \left( V_{F}\right) ^{n_{p,c}-1}-1\right\}
=D_{F}
\end{equation}

Finally, we can rewrite (\ref{__fe_P_V}, \ref{__fe_T_V}, \ref{__fe_Z}, \ref%
{__fe_Hp}) in the following form of the system of equations 
\begin{equation}
P_{2c}=\left( V_{F}\right) ^{n_{p,c}}P_{1c}  \label{__fe3_P_V}
\end{equation}%
\begin{equation}
T_{2c}Z_{2c}=\left( V_{F}\right) ^{n_{p,c}-1}T_{1c}Z_{1c}  \label{__fe3_T_V}
\end{equation}%
\begin{equation}
Z_{2c}=Z(T_{2c},P_{2c})  \label{__fe3_Z}
\end{equation}%
\begin{equation}
\frac{n_{p,c}}{n_{p,c}-1}\left\{ \left( V_{F}\right) ^{n_{p,c}-1}-1\right\}
=D_{F}  \label{__fe3_Hp}
\end{equation}

The last equation (\ref{__fe3_Hp}) depends only on $n_{p,c},$ the first
equation (\ref{__fe3_P_V}) depends only on $P_{2c}.$ $.$ So the variables
are almost separated.

\subsection{Solution Algorithm}

The solution algorithm to the system of equations for the polytropic
exponent and the outlet values of temperature, pressure, and compressibility
factor consists of the following steps:

1) From (\ref{__fe3_Hp}), find $n_{p,c}.$

2) From (\ref{__fe3_P_V}), find $P_{2c}.$

3) Substituting (\ref{__fe3_Z}) in (\ref{__fe3_T_V}), find $T_{2c}$ from
equation 
\begin{equation}
T_{2c}Z(T_{2c},P_{2c})=\left( V_{F}\right) ^{n_{p,c}-1}T_{1c}Z_{1c}
\label{__fe4_T2}
\end{equation}

On every step of the algorithm, there are used the solution methods for
one-dimensional nonlinear equation.

So finding the polytropic exponent allowed us to find the inlet pressure and
temperature. This makes it possible to find the polytropic efficiency by a
standard way as follows.

\subsection{Calculation of converted value of polytropic efficiency}

After the converted outlet state $\left( T,P\right) _{2,c}$ is determined
for the reference mixture composition $Mix_{r}$ and the reference inlet
state $\left( T,P\right) _{1,c}=\left( T,P\right) _{1,r}$ , the enthalpies $%
H_{1,c}$ and $H_{2,c}$ of the inlet and outlets states can be calculated due
to (\ref{f_h}). Ratio of the polytropic head with the enthalpy rise gives
the polytropic efficiency: 
\begin{equation}
E_{p,c}=\frac{H_{p,c}}{H_{2,c}-H_{1,c}}  \label{Ep_c}
\end{equation}

Then the converted required compression power in shaft is

\begin{equation}
W_{c}=\frac{1}{E_{p,c}}H_{p,c}m_{c}^{\prime }=\frac{1}{E_{p,c}}\frac{1}{%
A_{n_{p,c}}}\left( S_{c}^{A_{n_{p,c}}}-1\right) P_{1,_{c}}V_{1,_{c}}^{\prime
}
\end{equation}

We can conclude that all parameters of the operating point are converted.
Every such conversion of the operating point is defined after mechanical and
thermodynamic properties of both the process of compression of flow to be
converted and the reference state of the reference fluid at the inlet. So,
by construction, the conversions of the operating points are unique and
reversible. Hence they construct a group.

\section{Conversion of compressor characteristics}

Let be given any compressor characteristics for any mixture composition $Mix$
and any reference inlet state $\left( T,P\right) _{1}.$ How can they be
converted to another mixture composition $Mix^{\prime }$ and another
reference inlet state $\left( T^{\prime },P^{\prime }\right) _{1}?$ We show
now how to make a point-wise conversion.

Let us consider any compressor point corresponding to the given
characteristics as an actual operating point. So we redenote $Mix_{a}:=Mix,$ 
$T_{1,a}:=T_{1},$ $P_{1,a}:=P_{1}.$ Considering the another mixture
composition $Mix^{\prime }$ and another reference inlet state $\left(
T^{\prime },P^{\prime }\right) _{1}$ as the reference ones, we redenote $%
Mix_{r}:=Mix^{\prime },$ $T_{1,r}:=T_{1}^{\prime },$ $P_{1,r}:=P_{1}^{\prime
}.$ Let us implement the condition of full similarity of flow and according
procedure described above to any point of the given characteristics. Then we
convert the compressor characteristics as necessary.

\section{Numerical examples}

The results of conversion of the compressor characteristics are presented in
Fig. 1 -- 8 below. In these numerical examples, the thermodynamic properties
of real gases and gas mixtures were calculated by the use of
Soave-Redlich-Kwong equation of state. We have also performed calculations
for all numerical examples using the other equation of state, such as
Redlich-Kwong equation of state, Peng-Robinson equation of state, and some
others. Qualitatively, the results and figures remain valid.

From data given by vendor as diagramms of characteristics lines \thinspace $%
N=Const$ in coordinates $\left( V_{1}^{\prime },W\right) $ and $\left(
V_{1}^{\prime },P2\right) ,$ we calculated a variety of characteristics of a
compressor for 2 gases. One of these gases we used as an actual gas, another
one we used as a reference gas. More precisely, we used the second gas
composition as a normal case of the reference gas. For the actual gas, the
methane content is about 98\%. For the normal case of the reference gas, the
methane content is about 87\%. These characteristics will be called below as
the vendor characteristics given (i) for the actual gas and (ii) for the
normal case of the reference gas.

The conversion was made from the actual gas to the reference gas. The actual
inlet temperatures and pressures were kept identical to the reference inlet
temperature and pressure. Then we can compare the vendor characteristics for
the reference gas with the results of conversion. The comparison can be made
only for the converted parameters. If the condition of full similarity of
flow is used for conversion, then all parameters can be converted and
compared. Fig. 3 - 8 show that the comparable parameters and characteristics
lie one near another.

Otherwise, if only fan laws and similarity of flow at inlet (\ref%
{Ma1c_eq_Ma1a}, \ref{fi1c_eq_fi1a}, \ref{PSIc_eq_PSIa}) are used for
conversion, then such parameter as the polytropic efficiency cannot be
converted and compared. To make comparison possible, some people tried
before to use some additional suggestions. One of these is a suggestion that
the polytropic efficiency does not change and is invariant to the actual gas
and the actual temperature and pressure as well as to the reference gas and
the reference temperature and pressure

\begin{equation}
E_{p,c}=E_{p,a}  \label{id_Ep}
\end{equation}

Another suggestion is that there is a proportionality in some sense between
the isentropic exponent at inlet and the polytropic exponent of compression
process

\begin{equation}
\frac{A_{n,c}}{A_{n,a}}=\frac{A_{k1,r}}{A_{k1,a}}  \label{polyisentrop}
\end{equation}%
We call it as a polyisentropic proportionality.

If the polytropic exponent can be converted then the polytropic efficiency
can be converted as well. Indeed, the reduced oultet temperature $T_{2c}$
and pressure $P_{2c}$ can be found after (\ref{__fe3_P_V}, \ref{__fe3_T_V}, %
\ref{__fe3_Z}). Then the polytropic efficiency can be found after (\ref{Ep_c}%
).

Suppose, the polytropic exponent is converted under (\ref{Ma1c_eq_Ma1a}), (%
\ref{fi1c_eq_fi1a}), (\ref{PSIc_eq_PSIa})  and suggestion of polyisentropic
proportionality (\ref{polyisentrop}). Let us convert then the polytropic
efficiency. Its value contradicts to suggestion (\ref{id_Ep}) and vice
versa. Also if the polytropic exponent is converted under suggestion of full
similarity of flow then the value of polytropic efficiency contradicts to
suggestion (\ref{id_Ep}) and vice versa.

To compare the suggestion of full similarity of flow and the suggestion of
polyisentropic proportionality, let us take an extreme case. In the extreme
case of the reference gas, the methane content is about 77\% due to increase
of the isohexan content.

Fig. 1 shows the results of conversion under condition of polyisentropic
proportionality (\ref{polyisentrop}) and (\ref{Ma1c_eq_Ma1a}), (\ref%
{fi1c_eq_fi1a}), (\ref{PSIc_eq_PSIa}). The polytropic efficiency rises up to
120\% that is physically infeasible.

Fig. 2 shows the results of conversion under condition of full similarity of
flow (\ref{Ma1c_eq_Ma1a}), (\ref{fi1c_eq_fi1a}), (\ref{Ma2c_eq_Ma2a}), (\ref%
{fi2c_eq_fi2a}) and (\ref{PSIc_eq_PSIa}). The polytropic efficiency rises as
expected. Its values are still physically meaningful and acceptable.

\section{Conclusion}

We can conclude that the use of identity of Mach numbers at inlet $Ma_{1}$ (%
\ref{Ma1c_eq_Ma1a}), flow coefficients at inlet $\varphi _{1}$ (\ref%
{fi1c_eq_fi1a}) and process work coefficients $\psi $ (\ref{PSIc_eq_PSIa})
ensures conversion of inlet flow $V_{1}^{\prime }$, speed $N$ and polytropic
process work (head) $H_{p}$. Their converted three parameters $\left(
V_{1,c}^{\prime },H_{p,c},N_{c}\right) $ represent a point in the same
space, where a diagramm of compressor characteristics is often given by
manufacturer for the reference gas composition and the reference inlet
conditions. For any pair of inlet flow and speed $\left( V_{1,c}^{\prime
},N_{c}\right) ,$ a value of polytropic head $H_{p,e}$ can be read from
these manufacturer characteristics. The read value is called as expected
value. This allows a comparison between $H_{p,c},$ the operating value
converted to the reference conditions, and $H_{p,e},$ the expected value
given for these conditions by manufacturer or vendor.

No other value can be converted and compared, if it is used only identity of
Mach numbers at inlet (\ref{Ma1c_eq_Ma1a}), flow coefficients at inlet $%
\varphi _{1}$ (\ref{fi1c_eq_fi1a}) and process work coefficients $\psi $ (%
\ref{PSIc_eq_PSIa}). The performance monitoring can be constructed only
incompletely. Hence its availability and the use is limited.

The use of identity of Mach numbers and flow coefficients at inlet (\ref%
{Ma1c_eq_Ma1a}), (\ref{fi1c_eq_fi1a}), at outlet (\ref{Ma2c_eq_Ma2a}), (\ref%
{fi2c_eq_fi2a}) and process work coefficients $\psi $ (\ref{PSIc_eq_PSIa})
ensures conversion of inlet flow, speed, polytropic head, polytropic
exponent, outlet temperature, outlet pressure, and polytropic efficiency.
This allows a comparison of their operating values converted to the
reference conditions with their expected values given for these conditions
by manufacturer or vendor. The performance tests, comparison with guarantee
values, and performance monitoring can be constructed completely.

Furthermore, the conditions of full similarity of flow (\ref{Ma1c_eq_Ma1a}),
(\ref{fi1c_eq_fi1a}), (\ref{Ma2c_eq_Ma2a}), (\ref{fi2c_eq_fi2a}) and (\ref%
{PSIc_eq_PSIa}) ensures conversion of any compressor point and
characteristics from one gas composition taken for one inlet temperature and
pressure to another ones. These can be used by such applications as
simulation of operating points of compressors and drives; load sharing in
multi-machine configurations; load sharing in fluid networks; simulation of
fluid networks; optimization of fluid networks; optimal control. Moreover,
it simplifies the construction of these applications.

\bigskip

\textbf{Notation }

\textbf{Letters}

$A=$ area or cross section, $[m^{2}]$

$A_{k}=\frac{k-1}{k}$, $[-]$

$A_{n}=\frac{n-1}{n}$, $[-]$

$a=$ speed of sound, $[m/s]$

$b=$ outlet width of impeller, $[m]$

$C=$ conversion factor , $[-]$

$D=$ outlet diameter of impeller, $[m]$

$E=$ efficiency, $[-]$

$H=$ mass-specific enthalpy, $[kJ/kg]$

$H_{p}=$ specific polytropic work of compression i.e. specific polytropic
head, $[kJ/kg]$

$k=$ isentropic exponent, $[-]$

$m^{\prime }=$ mass flow, $[kg/s]$

$Ma=$ Mach number, $[-]$

$Mix=$ gas composition as a list of components, $[\%]$

$n=$ polytropic exponent i.e. polytropic volume exponent, $[-]$

$N=$ speed of rotation, $[1/s]$ (sometimes $[rps/rpm])$

$P=$ absolute pressure, $[bar]$

$R,R_{g}=$ specific gas constant, $[J/(kg\cdot K)]$

$R_{0}=$ universal gas constant, $[J/(kmol\cdot K)]$

$Re=$ Reynolds number, $[-]$

$S=$ pressure ratio, $[-]$

$T=$ absolute temperature, $[K]$

$u=$ velocity; linear tip speed, referred to $D$ of impeller, $[m/s]$

$v=$ velocity, $[m/s]$

$V=$ volume, $[m^{3}]$

$V^{\prime }=\dot{V}=$ volume flow, $[m^{3}/s]$

$W=$ power, $[W]$

$Y=$ specific work of compression i.e. specific head, $[kJ/kg]$

$Z=$ compressibility factor, $[-]$

$\mu =$ molar mass, $[kg/kmol]$

$\nu =$ kinematic viscosity, $[m^{2}/s]$

$\rho =$ density, $[kg/m^{3}]$

$\varphi =$ flow coefficient, $[-]$

$\psi =$ head coefficient i.e. process work coefficient, $[-]$

\bigskip

\textbf{Indexes}

$1=$ inlet

$2=$ outlet

$a=$ actual i.e. operating point of compressor

$e=$ expected value read from manufacturer characteristics

$c$ - converted point i.e. result of conversion of operating point of
compressor to the reference gas and the reference inlet conditions

$n=$ standard state i.e. standard condition for temperature and pressure
(STP)

$p=$ polytropic

$r$ - the reference gas composition and the reference state i.e. the
reference condition for temperature and pressure ($RTP$)

\bigskip

\textbf{Figures}

\bigskip

\bigskip



\begin{figure}[tbp]
\hspace{10.0cm} \centering
\includegraphics{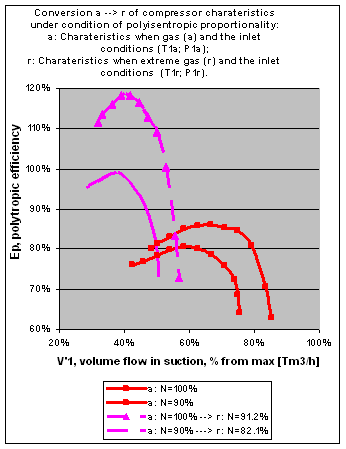}\newline
\caption{Efficiency in extreme case of the reference gas. Result is
infeasible.}
\label{fig:1}
\end{figure}

\bigskip

\begin{figure}[tbp]
\hspace{10.0cm} \centering
\includegraphics{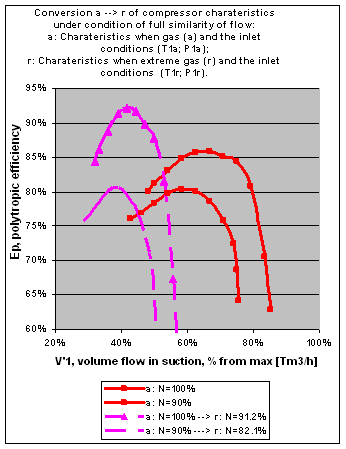}\newline
\caption{Efficiency in extreme case. Conversion is physically meaningful.}
\label{fig:2}
\end{figure}

\bigskip

\begin{figure}[tbp]
\hspace{10.0cm} \centering
\includegraphics{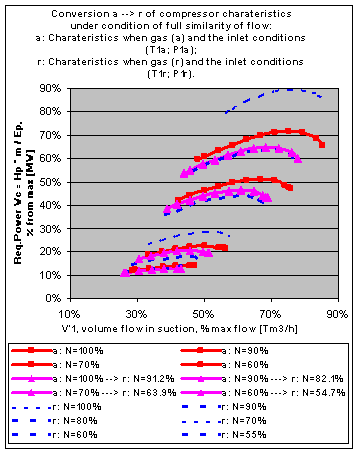}\newline
\caption{Compression power.}
\label{fig:3}
\end{figure}

\bigskip

\begin{figure}[tbp]
\hspace{10.0cm} \centering
\includegraphics{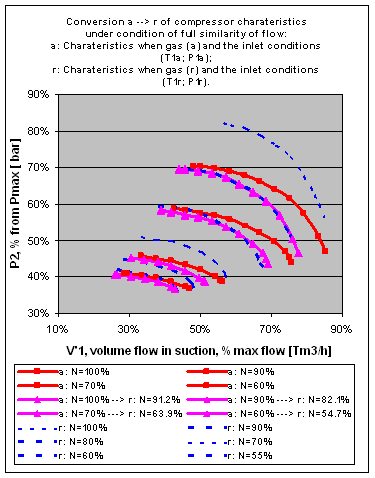}\newline
\caption{Outlet pressure.}
\label{fig:4}
\end{figure}

\bigskip

\begin{figure}[tbp]
\hspace{10.0cm} \centering
\includegraphics{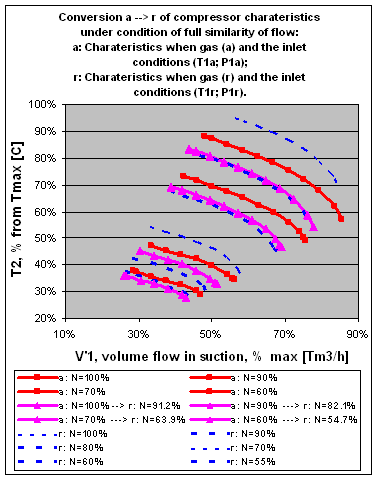}\newline
\caption{Outlet temperature.}
\label{fig:5}
\end{figure}

\bigskip

\begin{figure}[tbp]
\hspace{10.0cm} \centering
\includegraphics{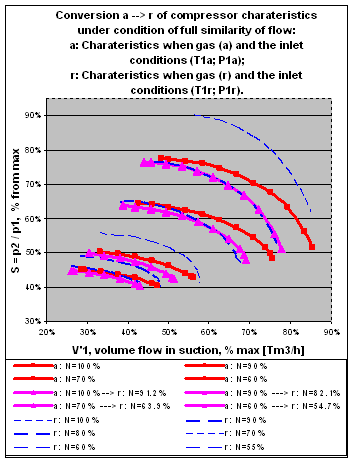}\newline
\caption{Pressure ratio.}
\label{fig:6}
\end{figure}

\bigskip

\begin{figure}[tbp]
\hspace{10.0cm} \centering
\includegraphics{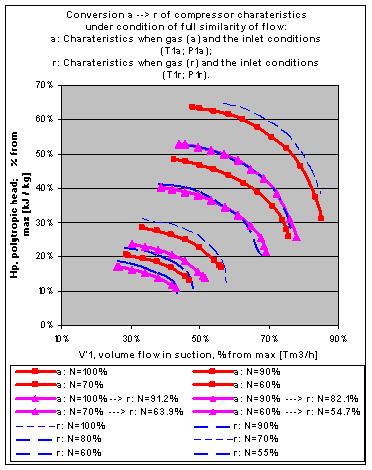}\newline
\caption{Polytropic head.}
\label{fig:7}
\end{figure}

\bigskip

\begin{figure}[tbp]
\hspace{10.0cm} \centering
\includegraphics{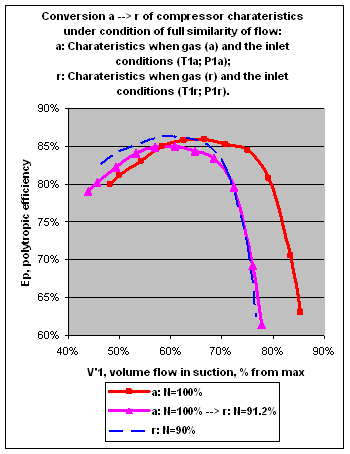}\newline
\caption{Polytropic efficiency.}
\label{fig:8}
\end{figure}

\bigskip

\end{document}